\def\be{\begin{equation}}
\def\ee{\end{equation}}
\def\beqn{\begin{eqnarray}}
\def\eeqn{\end{eqnarray}}

\def\ba{\begin{array}{c}}
\def\bat{\begin{array}{cc}}
\def\ea{\end{array}}
\def\bi{\begin{itemize}}
\def\ei{\end{itemize}}

\def\cR{{\cal R}}

\newcommand{\ta}[0]{\tilde \alpha}

\documentclass[3p,times,twocolumn]{elsarticle}

\usepackage{ecrc}
\usepackage{mathtools}
\usepackage{amsmath}
\usepackage{epstopdf}

\volume{00}

\firstpage{1}

\journalname{Nuclear Physics B Proceedings Supplement}

\runauth{}

\jid{nuphbp}

\jnltitlelogo{Nuclear Physics B Proceedings Supplement}

\usepackage{amssymb}

\usepackage[figuresright]{rotating}

\begin{document}

\begin{frontmatter}



\dochead{}

\title{Bounds on neutral and charged Higgs from the LHC}


\author{Victor Ilisie}

\address{Departament de F\'{\i}sica Te\`orica, IFIC,
        Universitat de Val\`encia -- CSIC,
        Apt. Correus 22085, E-46071 Val\`encia, Spain\\}

\begin{abstract}

After the discovery of a Standard Model-like boson with  mass of about 125 GeV the possibility of an enlarged scalar sector arises naturally. Here we present the current status of the phenomenology of the two-Higgs-doublet models with a special focus on the charged Higgs sector. If one considers a fermiophobic charged Higgs (it does not couple to fermions at tree level), all present experimental bounds are evaded trivially, therefore one needs to consider other decay and production channels. In this work we also present some of the interesting features of this specific scenario. 

\end{abstract}


\end{frontmatter}



\section{Introduction}

The discovery of a Higgs-like boson constitutes a great motivation in our quest for a deeper understanding of the scalar sector. Now that we have proven experimentally that this sector exists, one question that arises naturally is, are there more scalars?
The simplest extension of the SM, which has a richer scalar sector and that could give rise to new interesting phenomenology also in the flavour physics sector is the Two-Higgs-Doublet Model (2HDM) \cite{Ilisie:2014hea,Pich:2009sp,Celis:2013ixa,Ilisie:2013cxa,Celis:2013rcs,Jung:2013hka,Pich:2013vta,
Celis:2012dk,Jung:2012vu,Jung:2010ab,Jung:2010ik,Li:2014fea,
Gunion:1989we,Branco:2011iw,Barroso:2013zxa,Cheung:2013rva}. Next we shall present the phenomenology of the scalar sector of this model and see how the new LHC data together with the flavour constraints affects its parameter space.

\section{The Two-Higgs-Doublet Model}

The 2HDM extends the SM with a second scalar doublet of hypercharge $Y=\frac{1}{2}$.
The physical scalar spectrum contains five degrees of freedom: the two charged fields $H^\pm(x)$
and three neutral scalars $\varphi_i^0(x)=\{h(x),H(x),A(x)\}$, which are related with the $S_i$ fields
through an orthogonal transformation $\varphi^0_i(x)=\mathcal{R}_{ij} S_j(x)$. A detailed discussion is given in \cite{Ilisie:2014hea}. In this work we adopt the conventions\ $M_h \le M_H$\ and\
$ 0 \leq \ta \leq \pi$ therefore, $\sin{\ta}$ is positive.

\begin{figure}[!ht]
\centering
\includegraphics[scale=0.39]{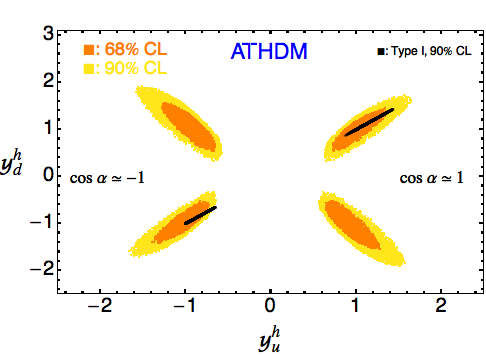}\\
\includegraphics[scale=0.39]{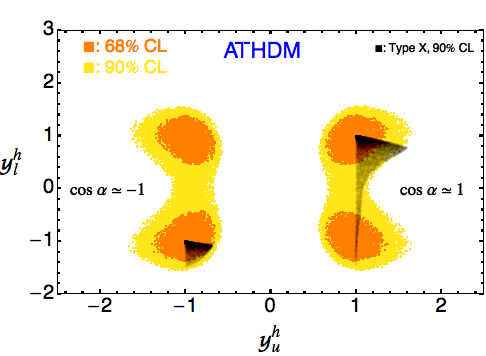}
\caption{\it One (orange, dark grey) and two (yellow, light grey) sigma allowed region for the parameter space of the A2HDM. The two sigma region for $\mathcal{Z}_2$ models are shown in black.}
\label{z2models}
\end{figure}

The most generic Yukawa Lagrangian with the SM fermionic content gives rise to dangerous tree level flavour changing neutral currents (FCNCs) which are phenomenologically suppressed. In order to get rid of them one usually imposes a discrete $\mathcal{Z}_2$ symmetry.  Here we consider the more generic approach given by the aligned two-Higgs-doublet model (A2HDM) \cite{Pich:2009sp}. In terms of the the mass-eigenstate fields the Yukawa Lagrangian reads
\begin{align}
\mathcal L_Y  =  & - \frac{\sqrt{2}}{v}\; H^+  \Big\{  \bar{u} \Big[ \varsigma_d\, V M_d \mathcal P_R - \varsigma_u\, M_u^\dagger V \mathcal P_L \Big]  d\,  \notag \\ 
 & \qquad \qquad\qquad\qquad+ \, \varsigma_l\, \bar{\nu} M_l \mathcal P_R l \Big\}
\nonumber \\
& -\,\frac{1}{v}\; \sum_{\varphi^0_i, f}\, y^{\varphi^0_i}_f\, \varphi^0_i  \; \left[\bar{f}\,  M_f \mathcal P_R  f\right]
\;  + \;\mathrm{h.c.} \notag
\end{align}
where $\mathcal P_{R,L}\equiv \frac{1\pm \gamma_5}{2}$ and where $\varsigma_f$ ($f=u,d,l$) are called the alignment parameters. These three parameters are independent, flavour universal, scalar basis independent and in general complex. Their phases introduce new sources of CP-violation and the usual models based on $\mathcal{Z}_2$ symmetries are recovered taking the appropriate limits \cite{Pich:2009sp}. The  couplings of the neutral scalar fields are given by:
\begin{align}
\label{yukascal}
 y_{d,l}^{\varphi^0_i} &= \cR_{i1} + (\cR_{i2} + i\,\cR_{i3})\,\varsigma_{d,l}   ,
\notag \\
 y_u^{\varphi^0_i} &= \cR_{i1} + (\cR_{i2} -i\,\cR_{i3}) \,\varsigma_{u}^* \, . 
\end{align}

\section{Phenomenology}

In the following we are going to consider the CP conserving limit of the 2HDM in the potential, as well as in the Yukawa sector.
For the first part we are going to present the impact of the current experimental data from the LHC on the parameter space related to the neutral scalar. As for the second part, we shall focus mainly on the charged Higgs sector, analyse the current bounds that we can draw from direct searches, and also present a few of the interesting features of the yet unexplored fermiophobic charged Higgs scenario.

\begin{figure}[!htb]
\centering
\includegraphics[scale=0.17]{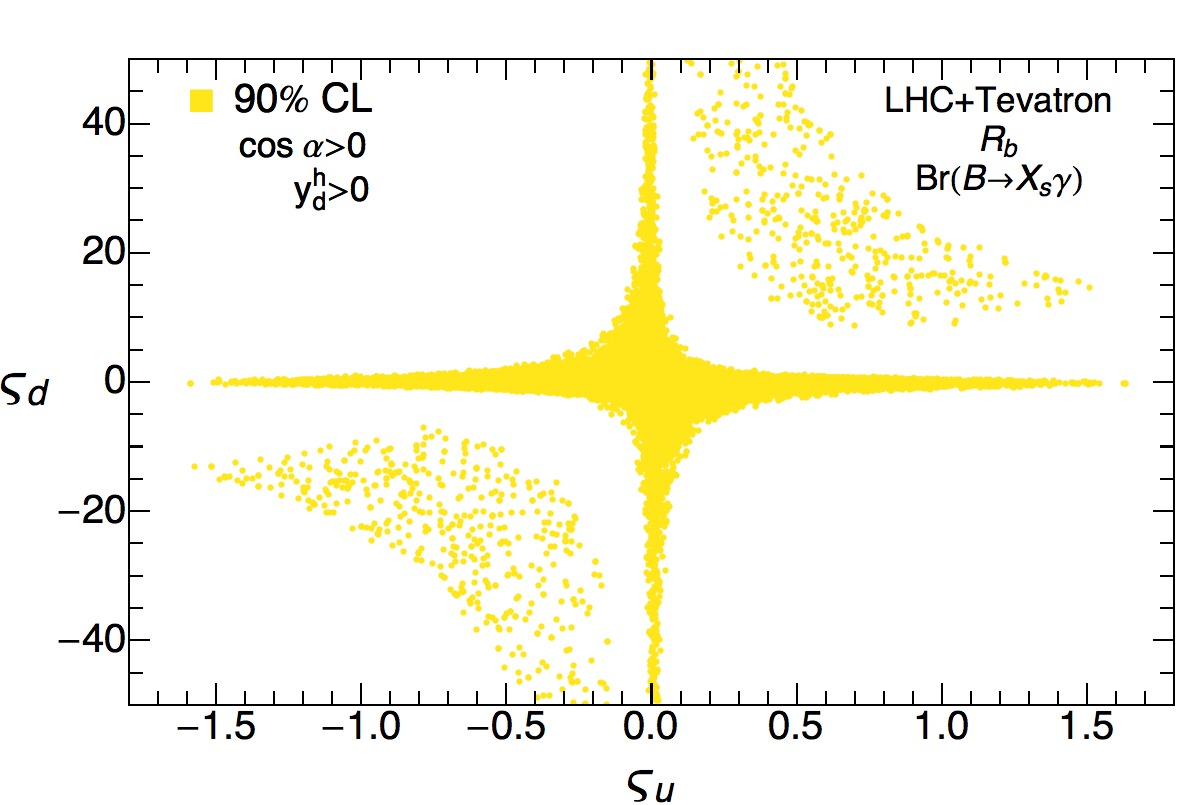} 
\caption{\it Allowed 90\% CL region (yellow-grey) in the plane $\varsigma_u - \varsigma_d$, from LHC and Tevatron data together with flavour constraints from $R_b$ and $\bar{B}\to X_s \gamma$.}
\label{combined}
\end{figure}

\subsection{Neutral sector}

The latest experimental data provided by the ATLAS \cite{Aad:2012tfa,Aad:2013wqa,ATLAS:2013sla} and CMS \cite{Chatrchyan:2012ufa,Chatrchyan:2013lba} collaborations from the LHC together with the latest combined results from Tevatron \cite{Aaltonen:2012qt,Aaltonen:2013kxa} are in good agreement with the SM hypothesis, but the experimental errors are still large. Assuming that the discovered scalar boson corresponds to the light CP-even Higgs, current data require its gauge coupling to be close to the SM one. A global fit \cite{Celis:2013ixa} to the LHC and Tevatron data gives
\begin{align}
|\cos\tilde\alpha| > 0.9 \, ,
\end{align}
at $68\%$ CL, where $g_{hVV}/g_{hVV}^{\mathrm{SM}} = \cos\tilde{\alpha}$. Some of the resulting constraints of the Yukawa couplings are shown in Fig.\ref{z2models}. The partial decay rates of the Higgs into a pair of fermions are not sensitive to the sign of the Yukawa coupling.
The only processes that are sensitive to the Yukawa relative signs are the loop induced ones {\it i.e.,} $gg\to h$, $h\to \gamma\gamma$, etc.  However, the dominating fermionic loop contribution is given by the top-quark. All other fermionic contributions are extremely suppressed, and this leads to the sign degeneracy of the Yukawa couplings we observe in Fig.\ref{z2models}. In the same figure, we can also observe the constraints on the models with a $\mathcal{Z}_2$ symmetry. The allowed regions (black) get considerably reduced in this case. This illustrates that there is a wider range of phenomenologically open possibilities of the A2HDM waiting to be explored.
In Fig.\ref{combined} we show the results of the combined fit from the LHC data together with flavour constraints \cite{Celis:2013ixa} in the  $\varsigma_u \, \varsigma_d$ plane. A very large part of this parameter space is excluded and these two parameters are found to be strongly correlated. This is mainly due to the $\bar{B}\to X_s \gamma$ radiative decay \cite{Jung:2010ab}. We shall see in the next section that these constraints are compatible and complementary with the ones given by direct charged Higgs searches.

\subsection{Charged sector}

\begin{figure}[!htb]
\centering
\includegraphics[scale=0.17]{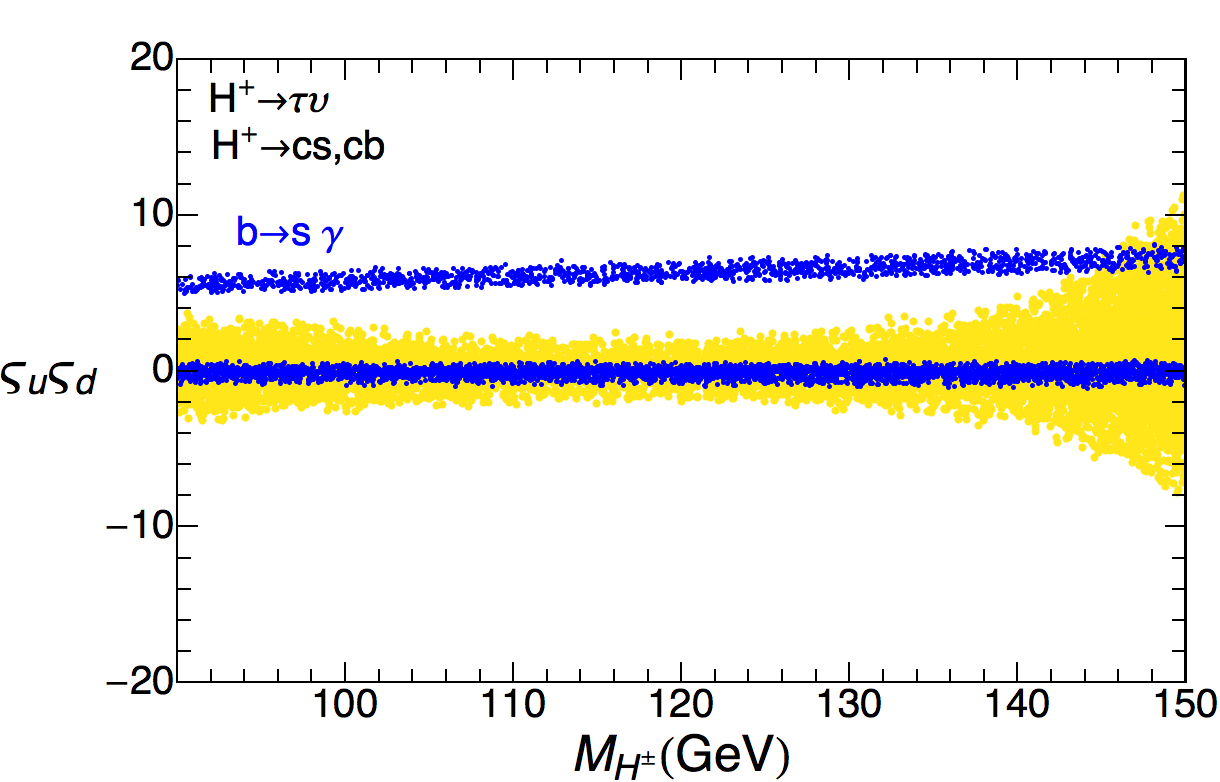} 
\caption{\it Allowed values for $\varsigma_u \, \varsigma_d$ as a function of the charged Higgs mass (yellow-light) obtained from the experimental 95\% CL upper bounds on the branching ratios. Allowed values for $\varsigma_u \, \varsigma_d$ from $\bar{B}\to X_s \gamma$ are shown in blue-dark.}
\label{directHpm}
\end{figure}

The LHC collaborations \cite{Aad:2012tj,Aad:2013hla,Chatrchyan:2012vca} have been performing searches for a charged scalar with yet negative results. Searches for a light charged Higgs mainly focus on the $t \to H^+ b$ production channel and fermionic decays like $H^+\to\tau^+\nu_\tau$ or $H^+\to c\bar{s}$. Therefore, from the searches with leptonic final states we can draw bounds on Br$(t\to H^+ b)\times$Br$(H^+\to\tau^+\nu_\tau)$.
For the di-quark final state searches the $H^+\to c\bar{s}$ is assumed to be the dominant decay rate and that is due to the CKM suppression $|V_{cb}| \ll |V_{cs}|$. However, in the A2HDM that is not always the case. If we write down the approximate formula
\begin{align}
\frac{\Gamma(H^+\to c\bar{b})}{\Gamma(H^+\to c\bar{s})} \approx \frac{|V_{cb}|^2 \; \big( |\varsigma_d|^2 \; m_b^2 + |\varsigma_u|^2 \; m_c^2  \big)}{|V_{cs}|^2 \; \big( |\varsigma_d|^2 \; m_s^2 + |\varsigma_u|^2 \; m_c^2  \big)} \notag 
\end{align}
we can observe that for $|\varsigma_d| \gg |\varsigma_u|$ the $H^+\to c\bar{b}$ can be important when compared to the other decay channels. Therefore the upper bounds on the di-quark final state must be interpreted as bounds on 
Br$(t\to H^+ b)\times$[Br$(H^+\to c\bar{s})$ + Br$(H^+\to c\bar{b})$]. In Fig.\ref{directHpm} we show the bounds on the A2HDM parameter space \cite{Celis:2013ixa} from direct searches of a light charged Higgs at the LHC together with flavour constraints from $\bar{B}\to X_s \gamma$. As it can be seen in the plot, part of the parameter space (the upper strip from $\bar{B}\to X_s \gamma$) is already excluded by the direct $H^\pm$ searches.

When $M_{H^\pm}>M_W + 2m_b$ there is a extra decay mode that can play a important role, and that is $H^+ \to t^* \bar{b} \to W^+ b\bar{b}$. This decay is normally very suppressed for large regions of the parameter space.
It has been previously analysed in the MSSM and 2HDMs with $\mathcal{Z}_2$ symmetries and it has been found to bring sizeable  contributions when $M_{H^\pm} \gtrsim 135-145$ GeV, depending on the model and on the value of $\tan\beta$.
In the A2HDM it can bring sizeable contributions, $\text{Br} \sim 10-20 \%$ already when $M_{H^\pm} \gtrsim 110$ GeV \cite{Celis:2013ixa}. If we re-analyse the previous constraints by adding this channel, the constraints stay roughly the same. However, there
are wide regions that partially overlap with the allowed parameter space region from direct searches. Therefore,
the experimental searches should be enlarged by also including this channel.

In Fig.\ref{directHpm} we observe that all experimental bounds are trivially satisfied if we set $\varsigma_f=0$. In this case, the charged Higgs becomes fermiophobic (does not couple to fermions at tree level). Therefore, in order to prove such a scenario,  other production channels and decay rates would be needed \citep{Ilisie:2014hea}.

In the last part of this section we are going to analyse the possibility of having a fermiophobic charged Higgs with a mass in the interval $M_{H^\pm} \in [M_W, \!\ M_W+M_Z]$. In this region, the only relevant decay rates are $H^+\to W^+\gamma$ and $H^+\to W^+\varphi_i^0$. For this last decay, the $W^+$ boson and the neutral scalar, cannot be both on-shell simultaneously for the whole considered kinematical range. Thus,
we shall consider three-body decays like $H^+\to W^+ f\bar{f}$ mediated by the neutral scalars $\varphi_i^0$
and $H^+\to \varphi_i^0 f_u\bar{f}_d$ mediated by a virtual $W^+$,
where $f_u\bar{f}_d$ stands for quark pairs $q_u\bar{q}_d$, or lepton-neutrino pairs $l^+\nu_l$.
The $H^+\to W^+ f\bar{f}$ final state is obviously dominated by the $b$-quark contribution, therefore we will neglect all other fermionic final states.
For the decay $H^+\to \varphi_i^0 f_u\bar{f}_d$, with an of-shell $W^+$ we are going to sum over all possible final states, quarks and leptons. We exclude the top quark, since this process is well below its production threshold. 

In the following we will consider the scalar boson $h$ with $M_h=125$ GeV as the light CP-even scalar of the model. 
A broad range of masses is allowed for the other scalars. Here we shall only present just a few of the interesting possibilities for the decay rates and production cross sections from \citep{Ilisie:2014hea}.

\begin{figure}[t]
\centering
\includegraphics[scale=0.49]{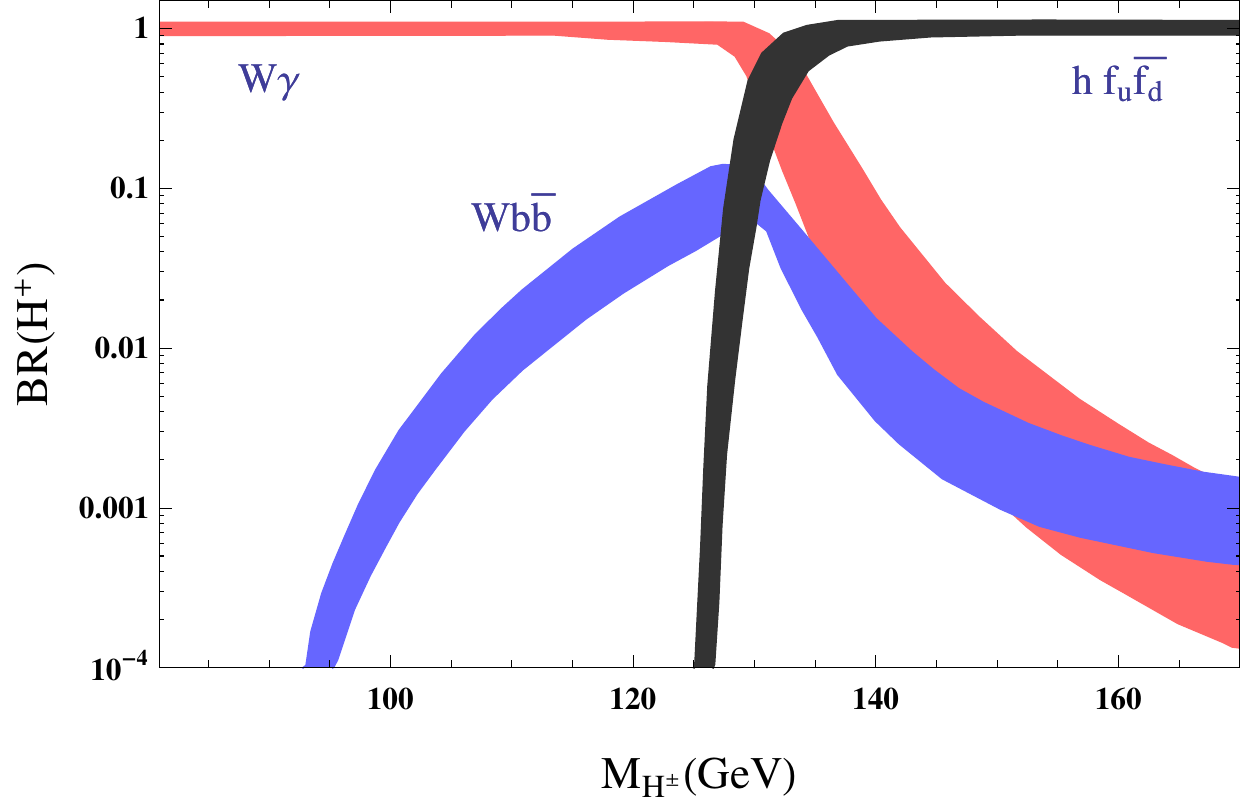}  \\[2ex]
\includegraphics[scale=0.49]{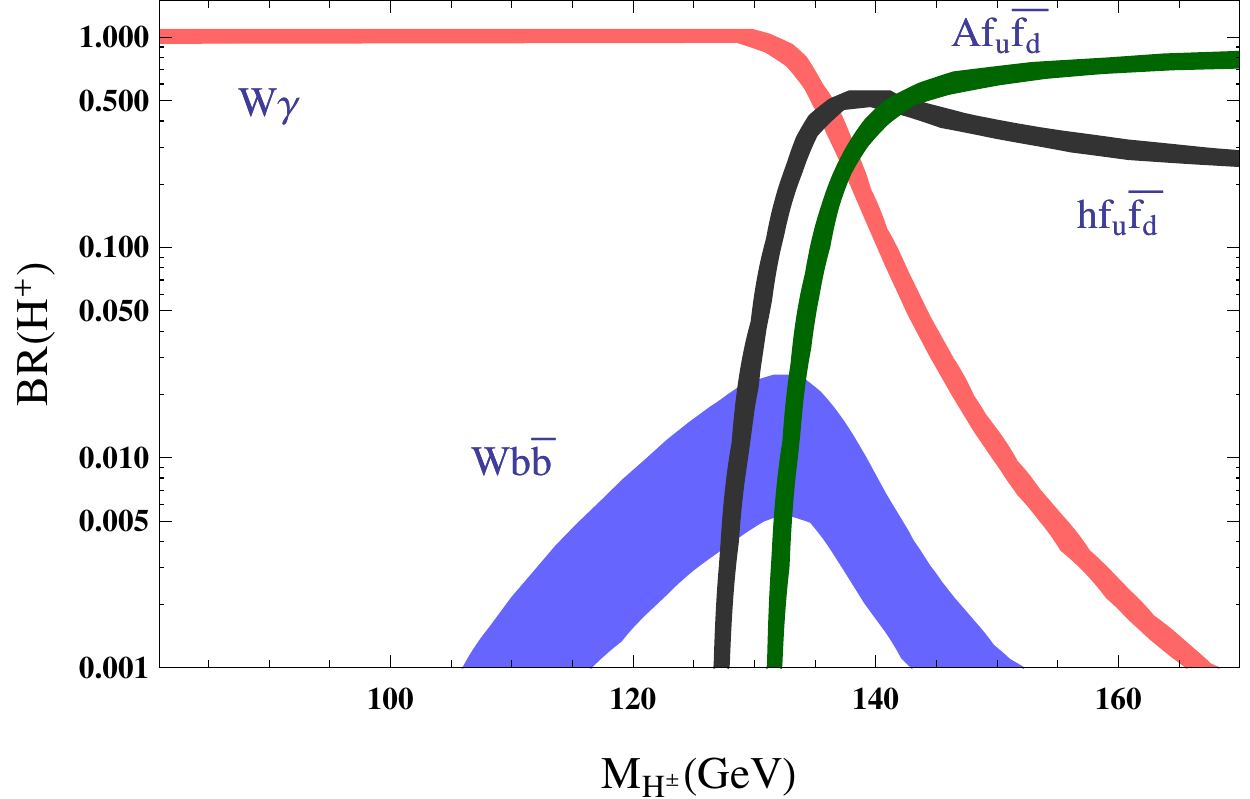}
\caption{\it Charged-Higgs branching ratios as functions of $M_{H^\pm} \in [M_W,M_W+M_Z]$, for $\cos\tilde\alpha = 0.9 $, $M_H \in [M_W+M_Z, \, 500\, \text{GeV}]$ and $\lambda_{h H^+ H^-} = \lambda_{H H^+ H^-} = 0$ (top), and for $\cos\tilde\alpha=0.9$, $M_A = $ 130 GeV and $M_H$ varied from $M_W + M_Z$ up to its permitted value by the oblique parameters (bottom).}
\label{Br1}
\end{figure}

For the first scenario, the masses of the remaining neutral scalars are considered to be greater than $M_W + M_Z$ so that decays of a charged Higgs into an on-shell $H$ or $A$ are kinematically forbidden. 
If we set $\cos\tilde\alpha = 0.9$ and the couplings $\lambda_{h H^+ H^-} = \lambda_{H H^+ H^-}= 0$, and vary the charged Higgs mass in the region $M_{H^\pm} \in [M_W,M_W+M_Z]$ and $M_H$ from $M_W + M_Z$ up to 500 GeV, we obtain the branching ratios shown in Fig.~\ref{Br1} (top). The decay channel $H^+\to W^+ \gamma$ dominates for $M_{H^{\pm}} \lesssim M_h$. When the charged Higgs is kinematically allowed to decay into an on-shell $h$, then $H^+\to hf_u\bar{f}_d$ rapidly becomes the dominant channel as $M_{H^\pm}$ grows. The remaining $H^+\to W^+ b\bar b$ branching ratio can be sizeable, raising up to the 10\% level. 

For the second scenario $M_H > M_W+M_Z$, as previously but this time we assume the CP-odd Higgs boson $A$ to have its mass below the $WZ$ threshold ($M_A < M_W+M_Z$).
The decay of the charged Higgs into an on-shell $A$ is now kinematically allowed. The same constraint as before is considered for the scalar mixing angle. 
If we take $\cos\tilde\alpha=0.9$, $\lambda_{hH^+H^-}=\lambda_{HH^+H^-}=1$ and $M_A=$ 130 GeV and vary $M_H$ from $M_W+M_Z$ up to its allowed upper bound from the oblique parameters (at 68\% CL), we obtain then the branching ratios in Fig.~\ref{Br1} (bottom). We observe that the decays into an on-shell $h$ and $A$ boson compete. However, the decay to $Af_u\bar{f}_d$ still dominates even if $M_h \sim M_A$ because of the relative suppression factor $\sin^2\tilde\alpha$ of the $hf_u\bar{f}_d$ width \citep{Ilisie:2014hea}. We observe that, the $H^+\to W^+b\bar{b}$ decay channel can bring relatively sizeable contributions in this case also.

\begin{figure}[t]
\centering
\includegraphics[scale=0.49]{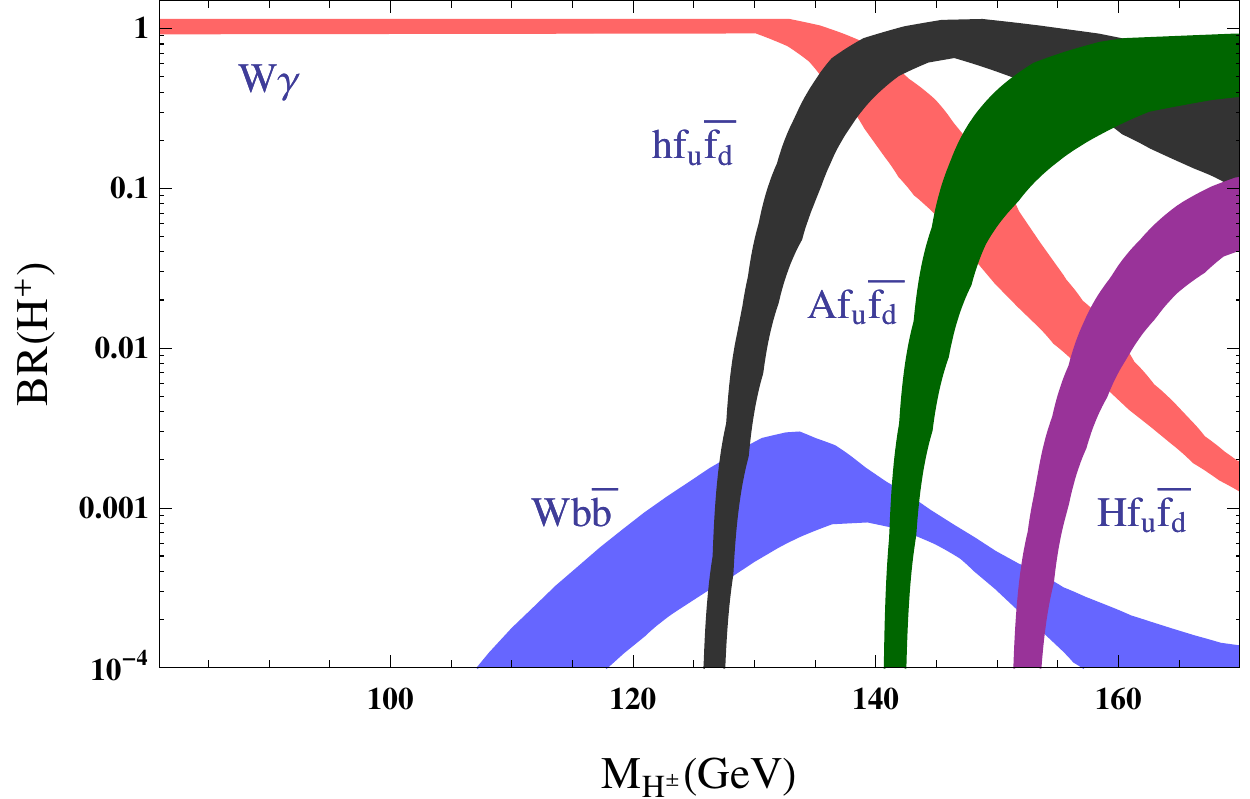}\\[2ex]
\includegraphics[scale=0.49]{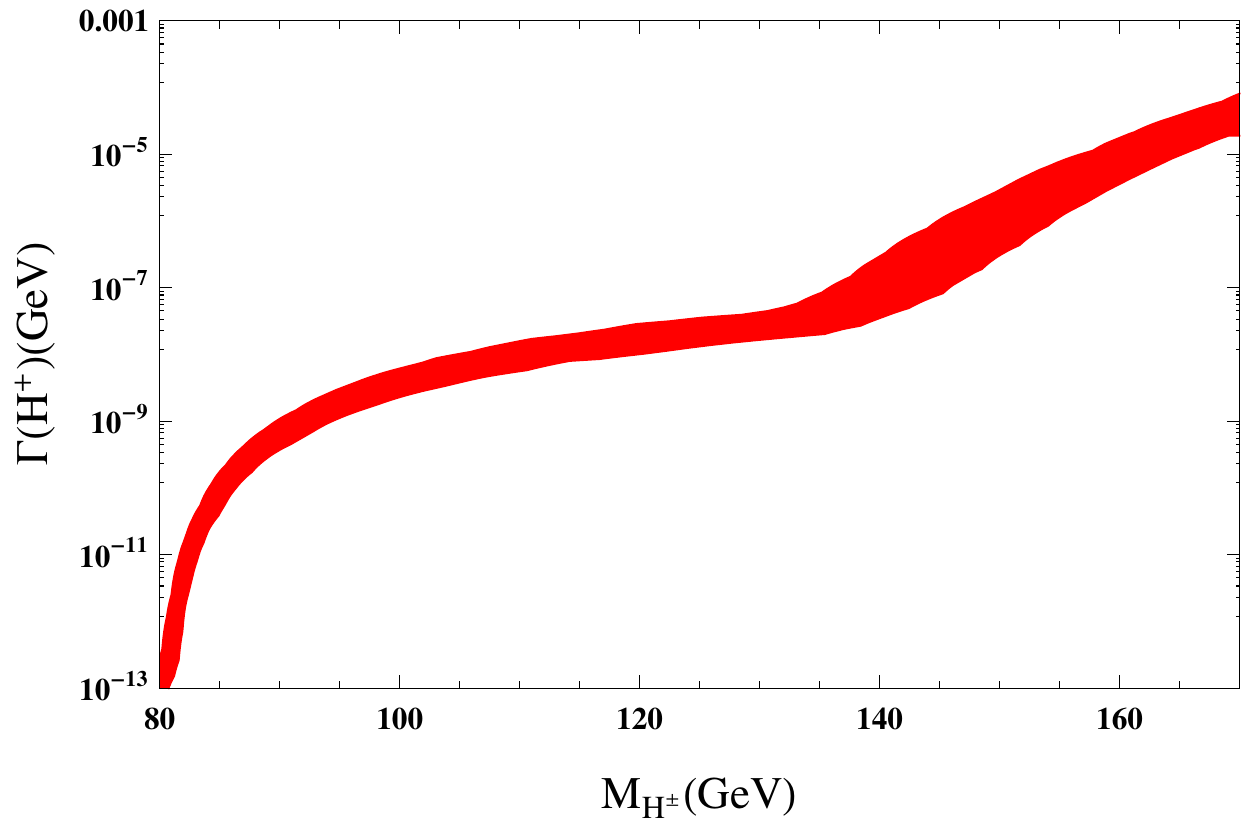}
\caption{\it Charged-Higgs branching ratios as functions of $M_{H^{\pm}}$, for $\lambda_{h H^+ H^-} = \lambda_{H H^+ H^-}= 1$, $\cos\tilde\alpha\in[0.9,\, 0.99]$,  $M_H = 140$ GeV, and $(M_H, M_A)=(150, 140)$ GeV (top). The total decay width as a function of $M_{H^\pm}$ for $\cos\tilde\alpha\in[0.9,\, 0.99]$,  $M_H = 140$ GeV, $M_A > M_W + M_Z $ is also shown (bottom).}
\label{Br2}
\end{figure}

For the third and last scenario, we allow the heavy CP-even Higgs boson $H$ to lie in the range $M_h < M_H < M_W+M_Z$. For the mass of the remaining CP-odd scalar we consider two possibilities: a) $M_A > M_W+M_Z$, so that the decay into an on-shell $A$ is forbidden and b) $M_A < M_H < M_W + M_Z$. For the last configuration, the $H^\pm$ boson could decay into any of the three neutral scalars. 
In Fig.~\ref{Br2} we show the $H^\pm$ branching ratios (top) for $(M_H, M_A)=(150, 140)$ GeV and the total decay width when $M_A > M_W+M_Z$ (bottom). In both cases we have set $\lambda_{h H^+ H^-} = \lambda_{H H^+ H^-}= 1$ and varied $\cos\tilde{\alpha}\in [0.9,\, 0.99]$.

As we have seen in the previously proposed scenarios, the configuration of the charged Higgs branching ratios depends very sensitively on the chosen parameters. However, we can draw some important conclusions. There are only a few decay channels to be analysed and the largest decay widths are the tree-level ones, corresponding to the on-shell production of scalar bosons. The number of decay channels decreases as the number of neutral scalar bosons that are heavier than the charged Higgs (i.e., $M_{\varphi_i^0}>M_{H^\pm}$) increases. The loop-induced $W\gamma$ decay rate can be sizeable below and close to the the on-shell production threshold of a scalar boson. Shortly after this threshold is reached, the $H^+\to W^+\gamma$ branching ratio rapidly decreases as $M_{H^{\pm}}$ grows.

\begin{figure}[t]
\centering
\includegraphics[scale=0.49]{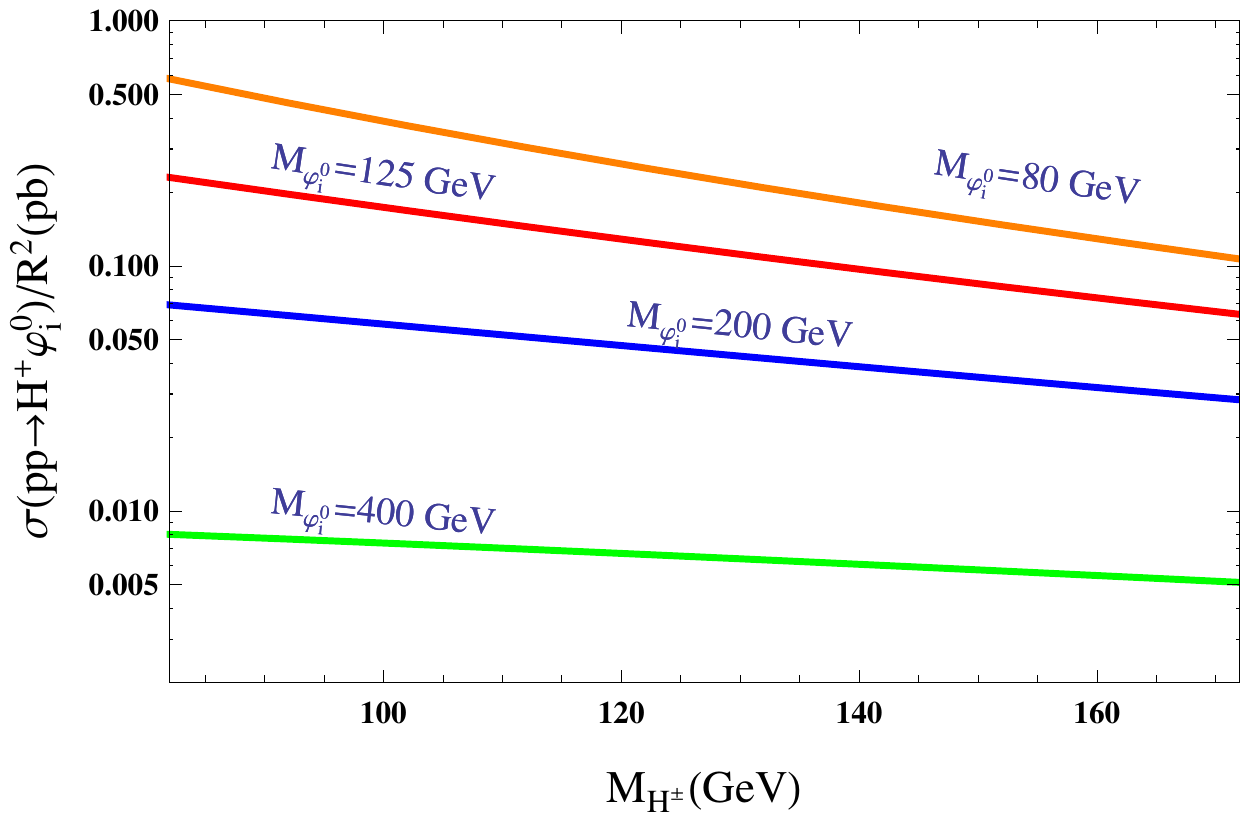} \\[2ex]
\includegraphics[scale=0.49]{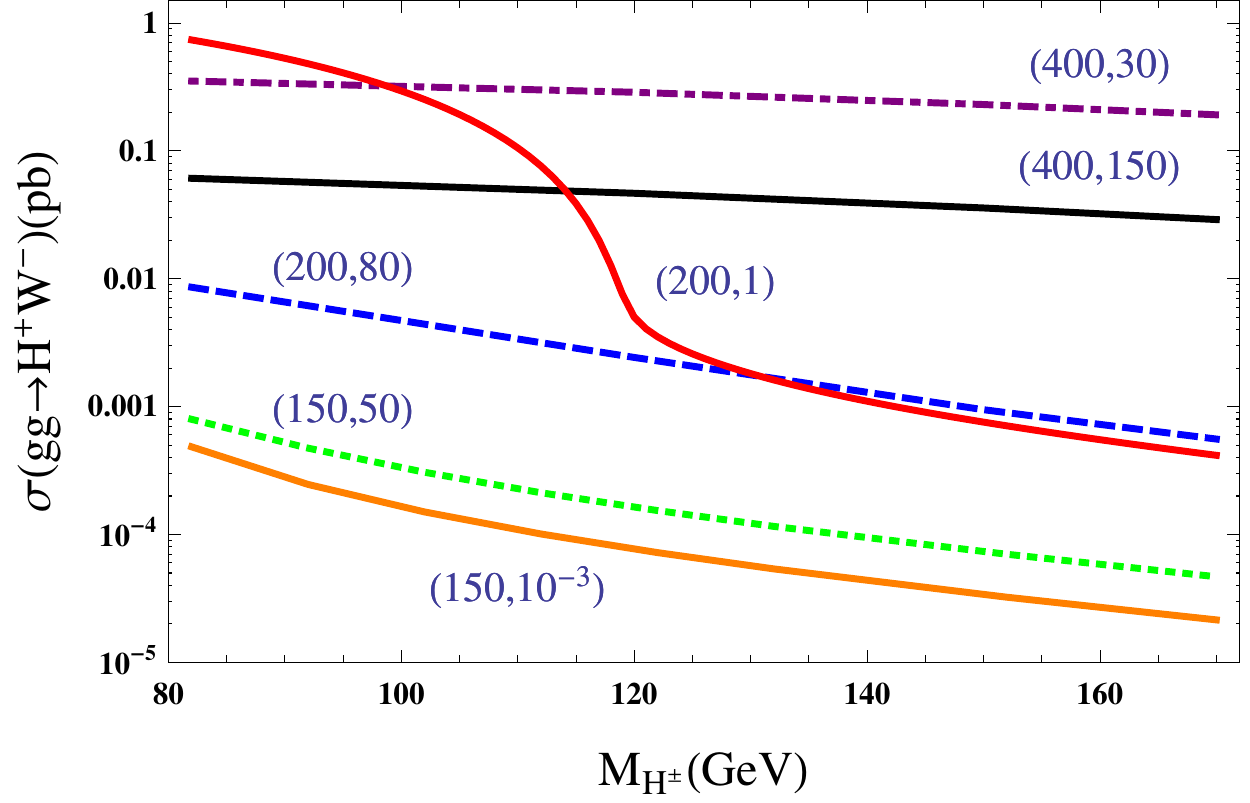}
\caption{\it LO production cross sections at $\sqrt{s}=14$~TeV as function of $M_{H^\pm}$. Top: $\sigma(pp\to H^+\varphi^0_i)/R^2$   for different values of
$M_{\varphi_i^0}$; bottom: $\sigma(pp\to H^+W^-)$ for $M_h=125$ GeV, $\cos\tilde\alpha=0.9$ and different values for the pair $(M_H,\; \Gamma_H)$ in GeV.}
\label{crossDY}
\end{figure}

In order to be able to experimentally probe the possibility of having a fermiophobic charged Higgs, one also needs an estimation of the production cross sections. There are two dominating channels, the associated production with a neutral scalar $q_u \bar{q}_d\to H^+ \varphi_i$ and the associated production with a $W$ boson $q\bar{q}/gg \to H^+W^-$.
The $q_u\bar{q}_d\to H^+\varphi_i^0$ production process is by far the most interesting channel, as it requires the least number of new parameters. As for the second process,
for typical LHC energies $gg\to H^+W^-$ production dominates over $q\bar q\to H^+W^-$. For hadronic center-of-mass energies $\sqrt{s}\sim 14$ TeV, the latter only corresponds at LO to a few percent of the total $pp\to H^+W^-$ cross section, so we can safely neglect it. The detailed expressions of the hadronic cross sections and the
QCD corrections are given in \citep{Ilisie:2014hea}.
Assuming the most general scalar potential, the LO cross section for $pp \to H^+ \varphi_i$ is proportional to the combination of rotation matrix elements $R^2\equiv(\mathcal{R}_{i2}^2+\mathcal{R}_{i3}^2)$. We take away the explicit dependence on the scalar-potential parameters, therefore $M_{\varphi_i^0}$ can be interpreted as the mass of any of the three neutral scalars of the theory. In Fig.~\ref{crossDY} (top) we plot the ratio
$\sigma(pp\to H^+\varphi_i^0)/R^2$ at $\sqrt{s}=14$~TeV, as a function of $M_{H^\pm}$ for different values of $M_{\varphi_i^0}$.
As we observe, the cross section reaches higher values for lower scalar masses. The interesting case is of course $M_{\varphi_i^0}$=125 GeV, since we already know that there is one scalar with that mass. If we consider $\varphi_i^0$ to be the light CP-even scalar of the theory, taking into account the $R^2=\sin^2{\tilde\alpha}$ suppression factor, this production channel can be experimentally quite challenging due to the small value of the cross section.
QCD corrections provide a mild enhancement, and the resulting QCD K factor $(K\equiv\sigma_{NLO}/\sigma_{LO})$ is typically around 1.2. 

For the remaining production channel $pp\to H^+W^-$, we have two alternative possibilities: we can either identify the 125 GeV boson with the lightest CP-even scalar $h$, or with the heaviest one $H$. For the last case, the cross section is extremely small, far below the experimental sensitivity in the near future. If we identify $M_h=125$ GeV with the mass of the light CP-even scalar, $H$ can be heavy enough to reach the on-shell region and it is necessary to regulate the propagator pole with its total decay width. 
Fig.~\ref{crossDY} (bottom) shows the predicted LO production cross sections at $\sqrt{s}=14$ TeV,
for representative values of $M_H$ and its total decay width $\Gamma_H$ covering the range of possibilities discussed in \citep{Ilisie:2014hea},
$(M_H,\; \Gamma_H)= (150,10^{-3})$, $(150,50)$, $(200,1)$, $(200,80)$, $(400,30)$, and $(400,150)$ GeV.
The cross section is very small when both CP-even scalars are off-shell. For $M_H = 150$ GeV, $\sigma(pp\to H^+W^-)$ is roughly smaller than $10^{-3}$ pb. With $M_H=200$ GeV and a large decay width $\Gamma_H=80$ GeV, the cross section stays below $10^{-2}$ pb; however,with a smaller width $\Gamma_H=1$ GeV, the cross section is enhanced by approximately two orders of magnitude
(three orders of magnitude with respect to the previous cases), in the region where $M_H$ is on-shell ($M_{H^\pm} \lesssim 120$ GeV).
The most interesting case is when $M_H=400$ GeV, because the cross section gets enhanced by the on-shell $H$ pole, reaching higher values around 0.1~pb.
For the other configurations both CP-even scalars are off-shell and the value of the cross section decreases by a few orders of magnitude. This results pretty challenging for the LHC, if not impossible. However, these small values could turn out to be measurable in the future if the LHC luminosity is increased.
The QCD K factor for this cross section is roughly around 1.9 for the whole range of the considered parameters, which is typical for gluon fusion processes.

\section{Conclusions}

The recent discovery of a Higgs-like boson has confirmed the existence of a scalar sector, which so far seems compatible with the SM predictions. The direct discovery of another scalar particle would bring priceless information on which type of extension of the scalar sector, is preferred by Nature. Here we have focused on the generic A2HDM scenario and we have presented some of the results of our previous analyses. We have shown the current bounds on the parameter space of the model that can be drawn from the LHC and other additional flavour constraints. We have mainly focused our attention on the charged Higgs, which is fundamental in searches for new physics. 

Besides the analysis of the {\it normal searches} (a charged Higgs that couples to fermions) at the LHC, we have presented an alternative scenario characterized by a fermiophobic charged Higgs.  It is a quite predictive case, since all Yukawa couplings are determined by the mixing among the neutral scalars.
The two most important production channels for a fermiophobic charged scalar have been investigated, including NLO QCD corrections.
Finally, if a fermiophobic charged Higgs is discovered in this
mass range, the precise value of its mass and its branching ratios would provide priceless information about the remaining parameters. The mean lifetime of a fermiophobic charged scalar is short, ranging from $10^{-11}$ to $10^{-23}$ s. For large values of this lifetime, the charged Higgs could produce a displaced vertex or even leave a trace in the detector. Therefore, the direct detection of such a particle looks very compelling at the LHC.

\section*{Acknowledgements}
I would like to thank the organizers of the ICHEP 2014 conference for giving me the opportunity of presenting this work. I would also like to thank A. Pich and A. Celis for their collaboration on this project. This work has been supported in part by the Spanish Government and ERDF funds from the EU Commission [Grants No. FPA2011-23778, No. CSD2007-00042 (Consolider Project CPAN)] and by Generalitat
Valenciana under Grant No. PROMETEOII/2013/007.





\nocite{*}
\bibliographystyle{elsarticle-num}
\bibliography{ilisie}

\begin{thebibliography}{10}
\expandafter\ifx\csname url\endcsname\relax
  \def\url#1{\texttt{#1}}\fi
\expandafter\ifx\csname urlprefix\endcsname\relax\def\urlprefix{URL }\fi
\expandafter\ifx\csname href\endcsname\relax
  \def\href#1#2{#2} \def\path#1{#1}\fi

\bibitem{Ilisie:2014hea}
V.~Ilisie, A.~Pich, {Low-mass fermiophobic charged Higgs phenomenology in
  two-Higgs-doublet models}, JHEP 1409 (2014) 089.
\newblock \href {http://arxiv.org/abs/1405.6639} {\path{arXiv:1405.6639}},
  \href {http://dx.doi.org/10.1007/JHEP09(2014)089}
  {\path{doi:10.1007/JHEP09(2014)089}}.

\bibitem{Pich:2009sp}
A.~Pich, P.~Tuzon, {Yukawa Alignment in the Two-Higgs-Doublet Model}, Phys.Rev.
  D80 (2009) 091702.
\newblock \href {http://arxiv.org/abs/0908.1554} {\path{arXiv:0908.1554}},
  \href {http://dx.doi.org/10.1103/PhysRevD.80.091702}
  {\path{doi:10.1103/PhysRevD.80.091702}}.

\bibitem{Celis:2013ixa}
A.~Celis, V.~Ilisie, A.~Pich, {Towards a general analysis of LHC data within
  two-Higgs-doublet models}, JHEP 1312 (2013) 095.
\newblock \href {http://arxiv.org/abs/1310.7941} {\path{arXiv:1310.7941}},
  \href {http://dx.doi.org/10.1007/JHEP12(2013)095}
  {\path{doi:10.1007/JHEP12(2013)095}}.

\bibitem{Ilisie:2013cxa}
V.~Ilisie, {Constraining the two-Higgs doublet models with the LHC data}, PoS
  EPS-HEP2013 (2013) 286.
\newblock \href {http://arxiv.org/abs/1310.0931} {\path{arXiv:1310.0931}}.

\bibitem{Celis:2013rcs}
A.~Celis, V.~Ilisie, A.~Pich, {LHC constraints on two-Higgs doublet models},
  JHEP 1307 (2013) 053.
\newblock \href {http://arxiv.org/abs/1302.4022} {\path{arXiv:1302.4022}},
  \href {http://dx.doi.org/10.1007/JHEP07(2013)053}
  {\path{doi:10.1007/JHEP07(2013)053}}.

\bibitem{Jung:2013hka}
M.~Jung, A.~Pich, {Electric Dipole Moments in Two-Higgs-Doublet Models}, JHEP
  1404 (2014) 076.
\newblock \href {http://arxiv.org/abs/1308.6283} {\path{arXiv:1308.6283}},
  \href {http://dx.doi.org/10.1007/JHEP04(2014)076}
  {\path{doi:10.1007/JHEP04(2014)076}}.

\bibitem{Pich:2013vta}
A.~Pich, {The Physics of the Higgs-like Boson}, EPJ Web Conf. 60 (2013) 02006.
\newblock \href {http://arxiv.org/abs/1307.7700} {\path{arXiv:1307.7700}},
  \href {http://dx.doi.org/10.1051/epjconf/20136002006}
  {\path{doi:10.1051/epjconf/20136002006}}.

\bibitem{Celis:2012dk}
A.~Celis, M.~Jung, X.-Q. Li, A.~Pich, {Sensitivity to charged scalars in
  $\boldsymbol{B\to D^{(*)}\tau\nu_\tau}$ and $\boldsymbol{B\to\tau\nu_\tau}$
  decays}, JHEP 1301 (2013) 054.
\newblock \href {http://arxiv.org/abs/1210.8443} {\path{arXiv:1210.8443}},
  \href {http://dx.doi.org/10.1007/JHEP01(2013)054}
  {\path{doi:10.1007/JHEP01(2013)054}}.

\bibitem{Jung:2012vu}
M.~Jung, X.-Q. Li, A.~Pich, {Exclusive radiative B-meson decays within the
  aligned two-Higgs-doublet model}, JHEP 1210 (2012) 063.
\newblock \href {http://arxiv.org/abs/1208.1251} {\path{arXiv:1208.1251}},
  \href {http://dx.doi.org/10.1007/JHEP10(2012)063}
  {\path{doi:10.1007/JHEP10(2012)063}}.

\bibitem{Jung:2010ab}
M.~Jung, A.~Pich, P.~Tuzon, {The B $\to$ Xs gamma Rate and CP Asymmetry within
  the Aligned Two-Higgs-Doublet Model}, Phys.Rev. D83 (2011) 074011.
\newblock \href {http://arxiv.org/abs/1011.5154} {\path{arXiv:1011.5154}},
  \href {http://dx.doi.org/10.1103/PhysRevD.83.074011}
  {\path{doi:10.1103/PhysRevD.83.074011}}.

\bibitem{Jung:2010ik}
M.~Jung, A.~Pich, P.~Tuzon, {Charged-Higgs phenomenology in the Aligned
  two-Higgs-doublet model}, JHEP 1011 (2010) 003.
\newblock \href {http://arxiv.org/abs/1006.0470} {\path{arXiv:1006.0470}},
  \href {http://dx.doi.org/10.1007/JHEP11(2010)003}
  {\path{doi:10.1007/JHEP11(2010)003}}.

\bibitem{Li:2014fea}
X.-Q. Li, J.~Lu, A.~Pich, {$B_{s,d}^0 \to \ell^+\ell^-$ Decays in the Aligned
  Two-Higgs-Doublet Model}, JHEP 1406 (2014) 022.
\newblock \href {http://arxiv.org/abs/1404.5865} {\path{arXiv:1404.5865}},
  \href {http://dx.doi.org/10.1007/JHEP06(2014)022}
  {\path{doi:10.1007/JHEP06(2014)022}}.

\bibitem{Gunion:1989we}
J.~F. Gunion, H.~E. Haber, G.~L. Kane, S.~Dawson, {The Higgs Hunter's Guide},
  Front.Phys. 80 (2000) 1--448.

\bibitem{Branco:2011iw}
G.~Branco, P.~Ferreira, L.~Lavoura, M.~Rebelo, M.~Sher, et~al., {Theory and
  phenomenology of two-Higgs-doublet models}, Phys.Rept. 516 (2012) 1--102.
\newblock \href {http://arxiv.org/abs/1106.0034} {\path{arXiv:1106.0034}},
  \href {http://dx.doi.org/10.1016/j.physrep.2012.02.002}
  {\path{doi:10.1016/j.physrep.2012.02.002}}.

\bibitem{Barroso:2013zxa}
A.~Barroso, P.~Ferreira, R.~Santos, M.~Sher, J.~P. Silva, {2HDM at the LHC -
  the story so far}\href {http://arxiv.org/abs/1304.5225}
  {\path{arXiv:1304.5225}}.

\bibitem{Cheung:2013rva}
K.~Cheung, J.~S. Lee, P.-Y. Tseng, {Higgcision in the Two-Higgs Doublet
  Models}, JHEP 1401 (2014) 085.
\newblock \href {http://arxiv.org/abs/1310.3937} {\path{arXiv:1310.3937}},
  \href {http://dx.doi.org/10.1007/JHEP01(2014)085}
  {\path{doi:10.1007/JHEP01(2014)085}}.

\bibitem{Aad:2012tfa}
G.~Aad, et~al., {Observation of a new particle in the search for the Standard
  Model Higgs boson with the ATLAS detector at the LHC}, Phys.Lett. B716 (2012)
  1--29.
\newblock \href {http://arxiv.org/abs/1207.7214} {\path{arXiv:1207.7214}},
  \href {http://dx.doi.org/10.1016/j.physletb.2012.08.020}
  {\path{doi:10.1016/j.physletb.2012.08.020}}.

\bibitem{Aad:2013wqa}
G.~Aad, et~al., {Measurements of Higgs boson production and couplings in
  diboson final states with the ATLAS detector at the LHC}, Phys.Lett. B726
  (2013) 88--119.
\newblock \href {http://arxiv.org/abs/1307.1427} {\path{arXiv:1307.1427}},
  \href {http://dx.doi.org/10.1016/j.physletb.2013.08.010}
  {\path{doi:10.1016/j.physletb.2013.08.010}}.

\bibitem{ATLAS:2013sla}
{Combined coupling measurements of the Higgs-like boson with the ATLAS detector
  using up to 25 fb$^{-1}$ of proton-proton collision data}.

\bibitem{Chatrchyan:2012ufa}
S.~Chatrchyan, et~al., {Observation of a new boson at a mass of 125 GeV with
  the CMS experiment at the LHC}, Phys.Lett. B716 (2012) 30--61.
\newblock \href {http://arxiv.org/abs/1207.7235} {\path{arXiv:1207.7235}},
  \href {http://dx.doi.org/10.1016/j.physletb.2012.08.021}
  {\path{doi:10.1016/j.physletb.2012.08.021}}.

\bibitem{Chatrchyan:2013lba}
S.~Chatrchyan, et~al., {Observation of a new boson with mass near 125 GeV in pp
  collisions at $\sqrt{s}$ = 7 and 8 TeV}, JHEP 1306 (2013) 081.
\newblock \href {http://arxiv.org/abs/1303.4571} {\path{arXiv:1303.4571}},
  \href {http://dx.doi.org/10.1007/JHEP06(2013)081}
  {\path{doi:10.1007/JHEP06(2013)081}}.

\bibitem{Aaltonen:2012qt}
T.~Aaltonen, et~al., {Evidence for a particle produced in association with weak
  bosons and decaying to a bottom-antibottom quark pair in Higgs boson searches
  at the Tevatron}, Phys.Rev.Lett. 109 (2012) 071804.
\newblock \href {http://arxiv.org/abs/1207.6436} {\path{arXiv:1207.6436}},
  \href {http://dx.doi.org/10.1103/PhysRevLett.109.071804}
  {\path{doi:10.1103/PhysRevLett.109.071804}}.

\bibitem{Aaltonen:2013kxa}
T.~Aaltonen, et~al., {Higgs Boson Studies at the Tevatron}, Phys.Rev. D88~(5)
  (2013) 052014.
\newblock \href {http://arxiv.org/abs/1303.6346} {\path{arXiv:1303.6346}},
  \href {http://dx.doi.org/10.1103/PhysRevD.88.052014}
  {\path{doi:10.1103/PhysRevD.88.052014}}.

\bibitem{Aad:2012tj}
G.~Aad, et~al., {Search for charged Higgs bosons decaying via $H^{+} \to \tau
  \nu$ in top quark pair events using $pp$ collision data at $\sqrt{s}=7$ TeV
  with the ATLAS detector}, JHEP 1206 (2012) 039.
\newblock \href {http://arxiv.org/abs/1204.2760} {\path{arXiv:1204.2760}},
  \href {http://dx.doi.org/10.1007/JHEP06(2012)039}
  {\path{doi:10.1007/JHEP06(2012)039}}.

\bibitem{Aad:2013hla}
G.~Aad, et~al., {Search for a light charged Higgs boson in the decay channel
  $H^+ \to c\bar{s}$ in $t\bar{t}$ events using pp collisions at $\sqrt{s}$ = 7
  TeV with the ATLAS detector}, Eur.Phys.J. C73 (2013) 2465.
\newblock \href {http://arxiv.org/abs/1302.3694} {\path{arXiv:1302.3694}},
  \href {http://dx.doi.org/10.1140/epjc/s10052-013-2465-z}
  {\path{doi:10.1140/epjc/s10052-013-2465-z}}.

\bibitem{Chatrchyan:2012vca}
S.~Chatrchyan, et~al., {Search for a light charged Higgs boson in top quark
  decays in $pp$ collisions at $\sqrt{s}=7$ TeV}, JHEP 1207 (2012) 143.
\newblock \href {http://arxiv.org/abs/1205.5736} {\path{arXiv:1205.5736}},
  \href {http://dx.doi.org/10.1007/JHEP07(2012)143}
  {\path{doi:10.1007/JHEP07(2012)143}}.

\end{thebibliography}







\end{document}